# Taylor-Couette Instability With a Continuous Spectrum

by
L.S. Yao and S. Ghosh Moulic[*]
Department of Mechanical and Aerospace Engineering
Arizona State University
Tempe, AZ 85287-6106

## Abstract

Nonlinear evolution of a continuous spectrum of unstable waves near the first bifurcation point in circular Couette flow has been investigated. The disturbance is represented by a Fourier integral over all possible axial wavenumbers, and an integrodifferential equation for the amplitude-density function of a continuous spectrum is derived. The equations describing the evolution of monochromatic waves and slowly-varying wave-packets of classical weakly nonlinear instability theories are shown to be special limiting cases. Numerical integration of the integrodifferential equation shows that the final equilibrium state depends on the initial disturbance, as observed experimentally, and it is not unique. In all cases, the final equilibrium state consists of a single dominant mode and its superharmonics of smaller-amplitudes. The predicted range of wavenumbers for stable supercritical Taylor vortices is found to be narrower than the span of the neutral curve from linear theory. Taylor-vortex flows with wavenumbers outside this range are found to be unstable and to decay, but to excite another wave inside the narrow band. This result is in agreement with the Eckhaus and Benjamin-Feir sideband instability. The results also show that a linearly stable long wave can excite a short unstable wave through nonlinear wave-interaction. An important implication of the existence of nonunique equilibrium states is that the torque induced by the fluid motion cannot be determined uniquely. The numerical results show that the *uncertainty*, associated with nonuniqueness, of using any accurately measured Taylor-vortex torque slightly above the first bifurcation point in engineering practice can be as large as ten percent.

---

[*] Present address, Department of Mechanical Engineering, Indian Institute of Technology, Bombay, India

The presence of multiple solutions at a fixed Reynolds number for a given geometry in Taylor-Couette flows has been known since Coles' monumental contribution in 1965. A theoretical confirmation has come only thirty years later. It is worthwhile to point out that the existence of multiple solutions, found by Coles, differs from current popular bifurcation theories. The current study indicates that the state of flows on a stable bifurcation branch can involve any wavenumber within a finite band and can not be determined uniquely. The multiple solutions in Coles' sense have also been found for mixed-convection flows (Yao & Ghosh Moulic 1993, 1994) besides the Taylor-Couette flows. We believe that the nonuniqueness of Coles sense, which complements the bifurcation theories, is a generic property for all fluid flows.

## 1. Introduction

Taylor (1923) showed that the flow between two concentric rotating cylinders can become unstable, resulting in a sequence of toroidal vortices spaced regularly along the axis of the cylinders. Due to its importance in the development of fluid dynamics, the instability of circular Couette flow has been the subject of many investigations. A considerable body of literature has accumulated on the linear-instability eigenvalue problem and the experimental measurements of the various transition boundaries. Much of the early work on the linear-instability problem is discussed by Chandrasekhar (1961). The development of finite-amplitude motions for Taylor numbers above the critical Taylor number was first studied by Stuart (1958), who used an energy-balance integral method to account for the distortion of the mean flow by the disturbance. Later, Davey (1962) used the formal expansion procedure of Stuart (1960) and Watson (1960) to compute the amplitude and form of the Taylor-vortex flow. Calculations based on the Stuart-Watson expansion can be carried out for any wavenumber lying in the unstable region of linear-instability theory near the neutral curve. These calculations predict supercritical equilibrium states. However, Kogelman & DiPrima (1970), following the method of Eckhaus (1965), showed that, in the limit as the Taylor number approaches its critical value, Taylor-vortex flows with wavenumbers outside a band of width $1/\sqrt{3}$ times the span of the neutral curve of linear theory and centered around the critical wavenumber $k_c$ of linear theory are unstable to axisymmetric disturbances. This instability is due to sideband perturbations which resonate with the first harmonic of the fundamental mode to mutually reinforce each other. The Eckhaus instability mechanism for two-dimensional flows that are periodic in one spatial dimension was shown to be related to the Benjamin-Feir (1967) instability mechanism of the two-dimensional Stokes water wave by Stuart & DiPrima (1978).

As the Taylor number is increased, the Taylor vortices become unstable at a second critical Taylor number, and a new flow pattern is established with azimuthally propagating waves superposed on the Taylor vortices. This wavy instability was first systematically studied by Coles (1965), who observed that the spatial structure of doubly periodic, Taylor-vortex flow (wavy Taylor-vortex flow) was not unique. He observed different axial and azimuthal wavenumbers at the same Taylor number. The nonuniqueness of wavenumber observed by Coles in doubly periodic flows was later shown to occur in the rotationally symmetric case by Snyder (1969) and Burkhalter & Koschmieder (1974). Snyder (1969) showed that, while the onset wavelength was unique, Taylor vortex flows of different wavenumbers could be obtained at the same value of the Taylor number by varying the initial conditions. He observed that there was a band of accessible wavenumbers narrower than the band that can grow according to linear theory. The extent of this band was found to be unsymmetrical about the onset wavenumber, with the increase occurring on the side of increasing wavenumber. Burkhalter & Koschmieder (1973) studied the variation of the wavelength of supercritical Taylor vortices experimentally. They found that singly periodic Taylor vortices maintained the critical wavelength over a wide range of Taylor numbers if the Taylor number was varied quasi-steadily and end effects were properly taken into account. In a later article, Burkhalter & Koschmieder (1974) reported results in which Taylor vortices were established through a sudden start of the inner cylinder. They found that a sudden start to a preset supercritical Taylor number gives the fluid a choice in the selection of the wavelength. Once established through a sudden start, a wavelength did not change its value if the Taylor number was maintained for a very long time or if the Taylor number was varied over larger intervals provided the flow was singly periodic.

In this investigation, the nonlinear interactions among Taylor vortices with different wavenumbers is studied using a perturbation method. The maximum amplification rate predicted by linear-instability theory is used as the expansion parameter. The disturbance is represented by a Fourier integral over all possible wavenumbers. Thus, the analysis is formulated with a continuous spectrum and therefore differs from previous theoretical studies. An integrodifferential equation describing the temporal evolution of the amplitude density of the wave components of a continuous spectrum is derived in §2. The evolution of monochromatic waves and slowly-varying wavepackets in classical weakly nonlinear instability theories are special limiting cases of the integrodifferential equation. However, the classical weakly nonlinear theories incorrectly assume a priori that the mean-flow distortion induced by nonlinear interactions is of a smaller order of magnitude compared to the amplitude of the dominant wave, as explained in the Appendix.

The choice of the small expansion parameter in the perturbation analysis in §2 differs from the small parameter used in the weakly nonlinear theory of monochromatic waves by Stuart (1960). The present ordering allows the possibility of nonlinear interactions among resonant triads of wavenumbers. Such nonlinear interactions among resonant triads are absent in monochromatic waves without considering its interaction with the leading-order deformation of the mean flow, but can occur in the case of a continuous spectrum of waves. The usual ordering employed in classical weakly nonlinear theories does not consider resonant triads, and leads to a result which violate the first thermodynamic law as explained in Appendix A. In the special limiting cases of monochromatic waves and slowly-varying wavepackets without considering the leading-order deformation of the mean flow, the final form of the amplitude equations obtained from the current formulation are the same as those in classical weakly nonlinear instability theories.

In §3, results are presented for the case when the outer cylinder is at rest and has a radius which is twice that of the inner cylinder. The integrodifferential equation describing the evolution of the amplitude density of the wave components has been solved with different initial conditions. In all cases, the final equilibrium state consisted of a single dominant mode together with its harmonics and an induced mean-flow distortion. However, the final dominant mode in the equilibrium state was found to depend on the waveform of the initial disturbance and the initial wave amplitudes; consequently, the equilibrium state is nonunique. These results are in agreement with the experimental observations of Snyder (1969) and Burkhalter & Koschmieder (1974). The range of wavenumbers of stable supercritical finite-amplitude Taylor vortices predicted by the present theory is found to be a subset of the linearly unstable band of wavenumbers. Taylor vortex flows with wavenumbers outside this subset, but within the linearly unstable band, are found to be unstable. This finding agrees with the Eckhaus and Benjamin-Feir sideband instability, and implies that a sideband instability is a consequence of nonlinear energy transfer among different interacting waves. Also, our results indicate that the selection of the equilibrium wave number is due to the intriguing nonlinear wave interactions and can be naturally determined.

The existence of multiple stable equilibrium states at the same Taylor number implies that the torque induced by the fluid motion cannot be determined uniquely. This nonuniqueness of the equilibrium state has also been observed in direct numerical simulations of mixed-convection flow in a heated vertical annulus using spectral methods employing a Fourier-Chebyshev expansion (Yao & Ghosh Moulic, 1994). We believe that multiple stable equilibrium is a generic property for all fluid flows. Consequently, any physical property transported by the fluid such as heat, salt etc. can at best be determined within a limit of uncertainty associated with nonuniqueness.

## 2. Analysis

Consider the flow between two infinitely long concentric circular cylinders with radii $r_1$ and $r_2$ ($> r_1$) rotating at angular velocities $\Omega_1$ and $\Omega_2$ respectively. The governing equations are the continuity and Navier-Stokes equations in cylindrical polar coordinates, $(r, \theta, z)$. Let $(u, v, w)$ denote the corresponding velocity components, p denotes the pressure, and t is the time. The flow is assumed to be rotationally symmetric so that u, v and w are independent of the azimuthal angle, $\theta$. All lengths have been scaled by the distance between the cylinders $d = r_2 - r_1$, the velocity components by $r_1 \Omega_1$, the time by $d^2/\nu$ and the pressure by $\rho r_1^2 \Omega_1^2$, where $\rho$ is the fluid density and $\nu$ is the kinematic viscosity. The Reynolds number is defined as $R = \Omega_1 r_1 d/\nu$.

The boundary conditions are $u = w = 0$, $v = 1$ when $r = r_i$ and $u = w = 0$, $v = \mu \left(1 + \dfrac{d}{r_1}\right)$ when $r = r_o$, where $\mu = \Omega_2/\Omega_1$ is the ratio of the angular speeds of the outer and inner cylinders respectively, $r_i = r_1/d$ and $r_o = r_1/d + 1$.

The basic state is steady laminar Couette flow, described by $u = w = 0$, $v = V(r) = A_0 r + \frac{B_0}{r}$, where $A_o = \frac{\mu - \eta^2}{\eta(1 + \eta)}$, $B_o = \frac{(1 - \mu)\eta}{(1 + \eta)(1 - \eta)^2}$ and $\eta = r_1/r_2$ is the radius ratio. The stability of this flow is studied by superposing a disturbance, which is unnecessarily to be small, on the basic state and writing the disturbed velocity field as $(u, v, w) = (u'(r, z, t), V(r) + v'(r, z, t), w'(r,z,t))$. In order to study the nonlinear interactions between the wave components of a continuous spectrum, the disturbance velocities are expressed as Fourier integrals over all possible axial wave numbers. Thus, the azimuthal velocity component is written as $v'(r, z, t) = \int_{-\infty}^{\infty} \hat{v}(k, r, t) e^{ikz} dk$. Similar expressions may be written for the other velocity components.

The linear instability of the basic state is studied by assuming the disturbance to be infinitesimal, and expressing the Fourier amplitudes of the disturbance quantities in separable form as

$$[\hat{u}(k,r,t), \hat{v}(k,r,t), \hat{w}(k,r,t), \hat{p}(k,r,t)] = [\tilde{u}(k,r), \tilde{v}(k,r), \tilde{w}(k,r), \tilde{P}(k,r)] e^{\sigma t},$$

where k is the axial wave-number and $\sigma$ is the amplification rate. The linearized disturbance equations may be written in terms of operators $L_1$, $L_2$ and $L_3$ as

$$D\tilde{u} + \frac{\tilde{u}}{r} + ik\tilde{w} = 0,$$

$$L_1(k, \sigma, \tilde{U}, \tilde{P}) = -\left(L - \frac{1}{r^2}\right)\tilde{u} - \frac{2RV(r)\tilde{v}}{r} + RD\tilde{P} + \sigma \tilde{u} = 0,$$

$$L_2(k, \sigma, \tilde{U}, \tilde{P}) = -\left(L - \frac{1}{r^2}\right)\tilde{v} + R\left(DV(r) + \frac{V(r)}{r}\right)\tilde{u} + \sigma \tilde{v} = 0, \quad (1)$$

$$L_3(k, \sigma, \tilde{U}, \tilde{P}) = -L\tilde{w} + ikR\tilde{P} + \sigma \tilde{w} = 0,$$

where $L \equiv D^2 + \frac{1}{r}D - k^2$, $D \equiv \frac{d}{dr}$ is the operator denoting differentiation with respect to r, and $\tilde{U} = [\tilde{u}, \tilde{v}, \tilde{w}]$. The boundary conditions are

$$\tilde{u} = \tilde{v} = \tilde{w} = 0 \text{ when } r = r_i \text{ and } r = r_o. \tag{2}$$

Equations (1) with the boundary conditions (2) form an eigenvalue problem for the amplification rate, $\sigma$, for each wave-number k. It is worth noting that for R not too large, the least stable eigenvalue $\sigma_1$ is real for all k, that is, the disturbed flow is non-oscillatory. The eigenfunctions $\tilde{u}$ and $\tilde{v}$ corresponding to the least stable eigenvalue are real, while the eigenfunction $\tilde{w}$ is purely imaginary. In this investigation, the eigenfunctions have been normalized so that $\int_{r_i}^{r_o} r[|\tilde{u}|^2 + |\tilde{v}|^2 + |\tilde{w}|^2] dr = 1$.

The nonlinear growth of the disturbances is studied by expanding the Fourier amplitudes in a perturbation series. The maximum amplification rate predicted by linear-instability theory is used as the expansion parameter $\varepsilon$. The amplification rate predicted by linear-instability theory for a wave of wavenumber k may be expressed in terms of $\varepsilon$ as $\sigma_1(k) = \varepsilon\, a_o(k)$, where $a_o(k)$ is a constant of order one and $\sigma_1(k)$ is the largest eigenvalue of the linear-instability operator for the mode with wave-number k. It may be noted that the small parameter $\varepsilon$ used in the present analysis differs from the small parameter used by in the weakly nonlinear theory of monochromatic waves Stuart (1960). The wave components of a continuous spectrum may exchange energy due to nonlinear interactions among resonant triads of wavenumbers. Such nonlinear interactions among resonant triads are absent in monochromatic waves. The present ordering allows the possibility of nonlinear interactions among resonant triads, and consequently differs from the usual ordering of classical weakly nonlinear theories of monochromatic waves. In the special limiting cases of monochromatic waves and slowly-varying wavepackets, the final form of the amplitude equations obtained from the current formulation is the same as those in the classical theories, as demonstrated later.

The expansion for the azimuthal disturbance velocity may be written as

$$\hat{v}(k,r,t) = \varepsilon\, \hat{v}_1(k,r,T_1,T_2) + \varepsilon^2\, \hat{v}_2(k,r,T_1,T_2) + \varepsilon^3\, \hat{v}_3(k,r,T_1,T_2) + \ldots \quad . \tag{3}$$

Here, $T_1 = \varepsilon t$ and $T_2 = \varepsilon^2 t$ are slow time scales. Expansions for the other dependent variables are given by similar expressions.

The first-order perturbation quantities are solutions of the linearized Navier-Stokes equations (1). They may be expressed in the form of separation of variables as

$$\hat{v}_1(k,r,T_1,T_2) = B(k,T_1,T_2)\, \tilde{v}_1(k,r), \tag{4}$$

where $\tilde{v}_1(k,r)$ is the eigenfunction of the linear-instability operator corresponding to the least stable eigenvalue, and $B(k, T_1, T_2)$ is a slowly-varying amplitude density function of order one.

The second order perturbation quantities are described by the equations

$$D\hat{u}_2 + \frac{\hat{u}_2}{r} + ik\hat{w}_2 = 0,$$

$$L_j(k,\sigma_1(k),\hat{U}_2,\hat{P}_2) = -\left[\frac{\partial B}{\partial T_1}(k,T_1,T_2) - a_o(k) B(k,T_1,T_2)\right] \tilde{U}_{1j}(k,r) \tag{5}$$

$$+ \int_{-\infty}^{\infty} B(k_1,T_1,T_2)\, B(k-k_1,T_1,T_2)\, F_j(k_1, k-k_1, r)\, dk_1,$$

where the inertial forcing terms $F_j$ depend on the linear-instability eigenfunctions (Ghash Moulic, 1993). It may be noted that the finite-amplitude disturbance for a selected wave number, k, has been expressed as a perturbation on a state of linear-stability analysis with a amplification rate, $\sigma_1(k)$ (Yao & Rogers, 1992). Thus, the operators on the left-hand sides of equations (5) are the same as those of linear instability theory, and a solution exists only if the right-hand side is orthogonal to the adjoint eigenfunction which satisfies the equation adjoint to (1). If the adjoint eigenfunctions are normalized so that $\int_{r_i}^{r_o} \left( u_1^+ \tilde{u}_1 + v_1^+ \tilde{v}_1 + w_1^+ \tilde{w}_1 \right) dr = 1$, application of the integrability condition results in the following integrodifferential equation describing the evolution of the amplitude density function on the $T_1$ time scale:

$$\frac{\partial B}{\partial T_1}(k, T_1, T_2) = a_o(k) B(k, T_1, T_2) + \int_{-\infty}^{\infty} b_o(k_1, k-k_1) B(k_1, T_1, T_2) B(k-k_1, T_1, T_2) dk_1, \qquad (6)$$

where the interacting constant $b_o(k_1, k-k_1)$ is given by

$$b_o(k_1, k-k_1) = \int_{r_i}^{r_o} \left[ u_1^+(k,r) F_1(k_1, k-k_1, r) \right.$$
$$\left. + v_1^+(k,r) F_2(k_1, k-k_1, r) + w_1^+(k,r) F_3(k_1, k-k_1, r) \right] dr. \qquad (7)$$

If equation (6) is satisfied, the solution to equations (5) may be expressed in the form

$$\hat{U}_2(k, r, T_1, T_2) = \int_{-\infty}^{\infty} B(k_1, T_1, T_2) B(k-k_1, T_1, T_2) \tilde{U}_2(k_1, k-k_1, r) dk_1, \qquad (8)$$

where the functions $\tilde{U}_2 = (\tilde{u}_2, \tilde{v}_2, \tilde{w}_2)$ are independent of the amplitude density function $B(k, T_1, T_2)$ and can be determined without solving the integrodifferential equation (6) (Ghash Moulic, 1993).

The disturbance velocity field at third order is described by the equations

$$D\hat{u}_3 + \frac{\hat{u}_3}{r} + ik\hat{w}_3 = 0 \,,$$

$$L_j(k, \sigma_1(k), \hat{U}_3, \hat{P}_3) = -\frac{\partial B}{\partial T_2}(k, T_1, T_2) \tilde{U}_{1j}(k,r) \qquad (9)$$

$$- \int_{-\infty}^{\infty} \left[ a_o(k_1) + a_o(k-k_1) - a_o(k) \right] B(k_1, T_1, T_2) B(k-k_1, T_1, T_2) \tilde{U}_{2j}(k_1, k-k_1, r) dk_1$$

$$+ \iint_{-\infty}^{\infty} B(k_1, T_1, T_2) B(k_2, T_1, T_2) B(k-k_1-k_2, T_1, T_2) G_j(k_1, k_2, k-k_1-k_2, r) dk_1 dk_2,$$

where the inertial forcing terms $G_j$ depend on the linear-instability eigenfunctions, $\tilde{U}_1$, and the functions $\tilde{U}_2$ determined at second order. As in the case of the second order equations, a solution to equation (9) exists only if an integrability condition is satisfied. Application of the Fredholm alternative theory yields a integrodifferential equation describing the evolution of the amplitude density function $B(k, T_1, T_2)$ on the $T_2$ - time scale:

$$\frac{\partial B}{\partial T_2}(k, T_1, T_2) = \int_{-\infty}^{\infty} b_1(k_1, k - k_1) B(k_1, T_1, T_2) B(k - k_1, T_1, T_2) dk_1$$

$$+ \int\int_{-\infty}^{\infty} c(k_1, k_2, k - k_1 - k_2) B(k_1, T_1, T_2) B(k_2, T_1, T_2) B(k - k_1 - k_2, T_1, T_2) dk_1 dk_2, \qquad (10)$$

where the interacting constants $b_1(k_1, k-k_1)$ and $c(k_1, k_2, k-k_1-k_2)$ are given by

$$b_1(k_1, k - k_1) = [a_o(k) - a_o(k_1) - a_o(k - k_1)] \int_{r_i}^{r_o} \left[ u_1^+(k, r) \tilde{u}_2(k_1, k - k_1, r) \right.$$

$$\left. + v_1^+(k, r) \tilde{v}_2(k_1, k - k_1, r) + w_1^+(k, r) \tilde{w}_2(k_1, k - k_1, r) \right] dr, \qquad (11)$$

$$c(k_1, k_2, k - k_1 - k_2) = \int_{r_i}^{r_o} \left[ u_1^+(k, r) G_1(k_1, k_2, k - k_1 - k_2, r) \right.$$

$$\left. + v_1^+(k, r) G_2(k_1, k_2, k - k_1 - k_2, r) + w_1^+(k, r) G_3(k_1, k_2, k - k_1 - k_2, r) \right] dr$$

The integrodifferential equations (6) and (10) obtained at second and third orders respectively may be combined to yield a single integrodifferential equation. In terms of a physical amplitude density function $A(k) = \varepsilon B(k)$, the integrodifferential equation may be written as

$$\frac{\partial}{\partial t} A(k, t) = a(k) A(k, t) + I_1(k, t) + I_2(k, t), \qquad (12)$$

where

$$I_1(k, t) = \int_{-\infty}^{\infty} b(k_1, k - k_1) A(k_1, t) A(k - k_1, t) dk_1, \qquad (13)$$

$$I_2(k, t) = \int\int_{-\infty}^{\infty} c(k_1, k_2, k - k_1 - k_2) A(k_1, t) A(k_2, t) A(k - k_1 - k_2, t) dk_1 dk_2, \qquad (14)$$

and the intrreracting constants a(k) and $b(k_1, k-k_1)$ are given by

$$a(k) = \varepsilon \ a_0(k), \quad b(k_1, k-k_1) = b_0(k_1, k-k_1) + \varepsilon \ b_1(k_1, k-k_1) \ . \tag{15}$$

It is worth noting that for R not too large, the interacting constants $a(k)$, $b(k_1, k-k_1)$ and $c(k_1, k_2, k-k_1-k_2)$ are real. The first interacting constant $a(k)$ represents the amplification rate of linear instability theory. The interaction integral $I_1(k,t)$ represents nonlinear interaction among all possible triads of wave-numbers $k_1$, $k-k_1$ and k, while the interaction integral $I_2(k,t)$ represents nonlinear interactions among all possible quartets of (primary) wavenumbers $k_1$, $k_2$, $k-k_1-k_2$ and k.

The integrodifferential equation (12) contains as a special case the equation describing the evolution of the amplitude of a discrete monochromatic wave. The amplitude density function for a discrete wave with wave-number $k_0$, without considering the leading-order deformation of the mean flow, has the form

$$A(k, t) = A_{k_o} \ \delta(k-k_0) + A_{k_o}^{*} \ \delta(k + k_0), \tag{16}$$

where $\delta(k)$ represents the Dirac delta function and the asterisk denotes complex conjugates. Substitution of (16) into equation (12) yields

$$\frac{dA_{k_0}}{dt} = a(k_0) A_{k_0} + a_1 \left|A_{k_0}\right|^2 A_{k_0} \ , \tag{17}$$

where

$$a_1 = c(k_0, -k_0, k_0) + c(-k_0, k_0, k_0) + c(k_0, k_0, -k_0) \tag{18}$$

is the second interacting constant and represents the resonant quatet of the disturbance wave and its complex conjugate. Equation (17) describes the evolution of the amplitude of a discrete monochromatic wave ( Stuart 1960 ). It is worth pointing out that there are no combinations of wavenumbers involving only the waves $\pm k_0$ which form a resonant triad. Thus, there is no contribution from the integral $I_1$ in the amplitude equation (17). The integrodifferential equation

(12) can also be reduced to a system of N ordinary differential equations describing the evolution of the amplitudes of a set of N discrete waves (Benney & Newell 1967). In Appendix A, we will show (17) violates the first law of thermodynamics.

Obviously, Eqn (12) describes the temporal evolution of the amplitude-density function in wave space. However, the integral formulation does include spatial variation of the velocity disturbances in physical space through the Fourier transformation, and is not restricted to periodic disturbances. In the special case of a wave packet near the critical wavenumber $k_c$ with a narrow spectral band width $\delta$, it becomes

$$v'(r, z, t) = \tilde{A}(k_c, z, t) \tilde{v}(k_c, r) e^{ik_c z} + c.c, \qquad (19)$$

where

$$\tilde{A}(k_c, z, t) = \int_{k_c - \delta}^{k_c + \delta} A(k, t) e^{i(k-k_c)z} \, dk = \varepsilon \int_{-\delta/\varepsilon}^{\delta/\varepsilon} A(k_c + \varepsilon K, t) e^{i\varepsilon K z} \, dK, \qquad (20)$$

is the envelope of the wave packet and c.c. denotes the complex conjugate. In order to derive an equation for the envelope of the wavepacket, we multiply equation (12) by $e^{i(k-k_c)z}$ and integrate with respect to k from $k_c - \delta$ to $k_c + \delta$. Using (20) and expanding the linear amplification rate in a Taylor series around $k = k_c$, it can be shown that in the limit as $\varepsilon \to 0$, equation (12) reduces to the equation for the spatio-temporal evolution of a wave packet derived by Newell & Whitehead (1969), Segel (1969) and Stewartson & Stuart (1971):

$$\frac{\partial \tilde{A}}{\partial t} = a(k_c) \tilde{A} + + a_1 |\tilde{A}|^2 \tilde{A} + a_2 \frac{\partial^2 \tilde{A}}{\partial z^2} \qquad (21)$$

where $a_1 = c(k_c, -k_c, k_c) + c(k_c, k_c, -k_c) + c(-k_c, k_c, k_c)$ and $a_2 = -\frac{1}{2} \frac{d^2 a}{dk^2}(k_c)$. It is worth noting that the nonlinear term in (28) which leads to an exchange of energy among the wave-components is identical to that for monochromatic waves. Equation (12), on the other hand, allows nonlinear exchange of energy among the wave components of a continous band of waves with widely differing wavenumbers. The advantages of this formulation are that it is simpler than previous theories and contains information on temporal and spatial evolution of nonlinear, interacting monochromatic waves, wave trains, and waves of a continuous spectrum.

### 3. Numerical solution and results.

The expansion of the azimuthal velocity component can be summarized as

$$v(r,z,t) = V(r) + \int_{-\infty}^{\infty} A(k,t)\tilde{v}_1(k,r)e^{ikz}dk$$
$$+ \int_{-\infty}^{\infty}\int_{-\infty}^{\infty} A(k_1,t)A(k-k_1,t)\tilde{v}_2(k_1,k-k_1,r)e^{ikz}dk_1dk$$
$$+ \int_{-\infty}^{\infty}\int_{-\infty}^{\infty}\int_{-\infty}^{\infty} A(k_1,t)A(k_2,t)A(k-k_1-k_2,t) \quad (22)$$
$$\tilde{v}_3(k_1,k_2,k-k_1-k_2,r)e^{ikz}dk_1dk_2dk$$
$$+....,$$

$A(k,t)$ becomes the final small parameter for our perturbation series. Results have been obtained for the case when the outer cylinder is at rest and has a radius which is twice that of the inner cylinder. The critical Reynolds number, $R_c$, predicted by linear instability theory for this case is 68.1, and the critical wavenumber at the onset of instability is $k_c = 3.16$. The Taylor number for this case is related to the Reynolds number by $T = \frac{64}{9}R^2$. In this investigation, we have obtained numerical results for a Reynolds number $R = 88.1$. Linear-instability analysis predicts that at this Reynolds number, circular Couette flow is unstable to disturbances with wavenumbers lying between 1.6 and 5.6. Thus, the range of wavenumbers permitted for supercritical Taylor vortices according to linear theory is $1.6 \leq k \leq 5.6$. The results indicate that the equilibrium state consists of monochromatic waves. The amplitude of each wave is $A_k = A \Delta k$. The values of $A_k$ differ slightly from the prediction by the monochromatic theory. The maximum value of $A_k$ for $R = 88.1$ is 0.1 which is small. We also calculated the kinetic energy associated with the base flow and the first-order perturbation terms for $k = 3$. The values are 0.1578 and 0.0096, respectively. Their ratio is about 6%. This indicates that the problem is weakly nonlinear. The results seem to demonstrate that the expansion for small $\varepsilon$, whose value is 8.34 for R = 88.1, may not necessarily require that $\varepsilon$ is smaller than one, but it should be smaller than a threshold value which is not known at present.

The integrodifferential equation (12) for the evolution of the amplitude density function A(k,t) was solved numerically using an implicit Euler scheme. The integrals $I_1$ and $I_2$ representing nonlinear interactions among resonant triads and quartets were discretized using the trapezoidal rule. The infinite range of integration was truncated to $-12 \leq k \leq 12$, which was found to be adequate. Most of the computations were done using a uniform mesh size $\Delta k = 0.25$ and a time step $\Delta t = 0.01$.

Figure 1 shows the equilibrium amplitudes, $A_k$, for supercritical Taylor vortices, obtained by solving the Landau equation (17) for discrete monochromatic waves. The maximum amplitude occurs at a wavenumber close to the minimum critical wavenumber $k_c = 3.16$ of linear theory. The equilibrium amplitude is zero for modes lying in the linearly stable region.

Figure 2(a) shows the results of a numerical simulation of a single dominant mode with a wavenumber k = 3, obtained by solving equation (12) for the evolution of the amplitude density function starting with initial condition A(k,0) = 0.125 for the mode k = 3 and zero for the other modes. The mode k = 3 is linearly unstable and grows by obtaining energy from the mean flow. As its amplitude increases, nonlinear effects become important and alter the linear growth rate. The figure indicates that nonlinear interactions lead to the generation of the second harmonic (k = 6), and distort the mean flow (k = 0). This may be explained by referring to Figure 2(b) in which the interaction integrals $I_1$ and $I_2$ in the integrodifferential equation (19) have been plotted as a function of time. Figure 2(b) shows that $I_1$ is negative for the mode k = 3, and positive for the modes k = 6 and k = 0. Thus, energy is transferred nonlinearly from the mode k = 3 to its harmonic k = 6, and the mean flow (k = 0) through a three-wave resonance. The interaction integral $I_2$ is negative for all the modes. Thus, the four-wave resonance mechanism has a stabilizing influence on all the modes. The mode k = 3 grows and eventually reaches equilibrium together with its harmonic (k=6) and the mean flow (k=0). In summary, the energy is transferred linearly from the mean flow to the disturbance (k=3) and tranfserred nonlinearly from the wave (k=3) to the mean flow (k=0) and its harmonics (k=6...). The deformation of mean flows is at same order of magnitude of the dominant wave, and is not one order of magnitude smaller than the dominant wave as assumed by the classical weakly nonlinear instability theories (see Appendix A). It is worth noting that the function A(k,t) plotted in Figure 2(a) is an amplitude density function, and the corresponding wave-amplitudes of the modes are given by the area $\overline{A}_k = A(k)\Delta k$. The equilibrium amplitude $\overline{A}_k$ for the mode k = 3 differs slightly from the equilibrium amplitude $A_k$ predicted by the Landau equation (17) for a discrete monochromatic wave near the onset of the instability. This is due to the terms of the resonant triads and quartets missing from the Landau equation as explained in the Appendix A. An additional computation with a smaller size of $\Delta k$ =0.2 predictes the same value of $\overline{A}_k$. This indicates that the resolution used in our computations is adequate, although the amplitude density function A(k,t) for a single dominant mode depends on the mesh size $\Delta k$ used in the calculation. The dependence of the amplitude density function on the mesh size $\Delta k$ for a single dominant mode is expected since in the limit as $\Delta k$ approaches zero, the amplitude density function should approach a delta function. Thus, if $\Delta k$ is reduced, the amplitude density A(k) should increase, the area $A(k)\Delta k$ being conserved.

The results of a numerical simulation starting with a single dominant mode with wavenumber k = 2 at time t = 0 is presented in Figure 3(a). The mode k = 2 is linearly unstable, and the weakly nonlinear theory of discrete monochromatic waves predicts a supercritical equilibrium amplitude for this mode. However, as indicated in the figure, the mode k = 2 decays to zero, while its harmonic k = 4, excited through nonlinear interactions, grows and reaches a finite-amplitude equilibrium state. This indicates that the equilibrium state predicted by the weakly nonlinear theory of discrete monochromatic waves for the mode k = 2 is unstable. This result is in agreement with the Eckhaus and Benjamin-Feir sideband instability. This implies that a sideband instability is a consequence of nonlinear energy transfer among different interacting waves. Figure 3(b) shows that the mode k = 4 receives energy through the three-wave resonance mechanism, as the interaction integral $I_1$ representing nonlinear interactions among triads of wavenumbers is negative for the mode k = 2, and positive for the mode k = 4. This indicates that the instability is triggered by a three-wave resonance mechanism. Then, the mode k=4 receives the energy linearly from the mean flow, and transfers energy to the mean flow and its harmonics nonlinearly.

Figure 4 is a summary of numerical simulations in which a single mode was given an initial amplitude density of 0.1 at time t = 0 and the other modes were given a small initial amplitude density of $10^{-5}$. The equilibrium wave amplitude $A_k = A(k)\Delta k$ has been plotted in Figure 4, rather than the equilibrium amplitude density function A(k). The solid lines in the middle of the figure refer to modes which reach a finite-amplitude equilibrium state, while the dotted lines to the left and right indicate modes which decay to zero, their energy being transferred to other modes. The final dominant wavenumber in the case of initial disturbances with wavenumbers k = 1.75, 2, 2.25 and 2.5, indicated by the dotted lines on the left of the figure, was their first harmonic k = 3.5, 4, 4.5 and 5, respectively. These are generated through nonlinear interaction. The final dominant wavenumber in the case of initial disturbances with wavenumbers k = 5.25 and 5.5, indicated by the dotted lines on the right of the figure, was k = 3.25. The numerical simulations were repeated with initial conditions consisting of a larger uniform broad-band spectrum with an amplitude density of 0.01 superimposed on the single dominant initial mode which was given an amplitude density of 0.1. The final equilibrium state remained unchanged in all cases except for the modes with wavenumbers k = 2.5 and 5. In the latter cases, the final dominant mode was found to be k = 3. Numerical simulations starting with initial conditions consisting of a single mode at time t = 0 and zero background noise for the other modes yielded the same equilibrium state as with a small uniform background noise with initial amplitude density of $10^{-5}$ in all cases except for the modes with wavenumbers k = 5.25 and 5.5, indicated by the dotted lines on the right end of the figure. In the absence of other modes at time t = 0, the modes k = 5.25 and 5.5 remained the dominant ones in the final equilibrium state. This is because nonlinear interactions can only generate higher harmonics. Thus, energy can be transferred to subharmonic modes through nonlinear interactions only if these modes are present at time t = 0, since they cannot be created through the nonlinear interactions. However, as indicated by Figure 4, the presence of very low levels of noise in the subharmonic modes can destabilize Taylor vortex flows with wavenumbers k = 5.25 and 5.5. Consequently, supercritical Taylor vortex flows with these wavenumbers are unlikely to be observed experimentally at the Reynolds number used in our computations. Figure 4 indicates that the range of wavenumbers for which our theory predicts *stable* supercritical Taylor vortices is a subset of the linearly unstable range of wavenumbers.

Figure 5(a) shows the results of a numerical simulation in which two modes k = 3.25 and k = 3.75 were given initial amplitudes of 0.2 at time t = 0. The figure shows that nonlinear interactions generate the harmonics k = 6.5, 7 and 7.5. The mode k = 3.75 decays as energy is transferred to the harmonics and the mean flow, while the mode k = 3.25 grows as it obtains energy from the mean flow, and eventually reaches a supercritical finite-amplitude equilbrium state. Figure 5(b) shows the results of a numerical simulation in which the mode k = 3.25 was given an initial amplitude of 0.1 and the mode k = 3.75 was given an initial amplitude of 0.2. In this case, the mode k = 3.75 remains the dominant one in the final equilibrium state, while the mode k = 3.25 decays to zero. This demonstrates that an equilibrium state can shift to a new equilibrium state by a finite disturbance. This property can have significant practical applications.

Numerical simulations starting with a uniform broad-band spectrum at time t = 0 result in a final equilibrium state consisting of a single dominant wave and its harmonics, and an induced mean flow distortion. For initial amplitudes smaller than 0.05, the final equilibrium wave is k = 3.5. This is because the nonlinear terms are initially small, and hence the initial growth of the disturbances is according to linear theory. The mode k = 3.5 has the largest linear growth rate at the Reynolds number used in this computation. Thus, this mode grows faster than the other modes, and remains the dominant mode in the final equilibrium state. With initial amplitudes larger than 0.05, nonlinear effects become important at an earlier stage in the evolution of the waves, and the dominant wave in the final equilibrium state is k = 3.25, which is the one closest to the minimum critical wavenumber, 3.16, at the onset of instability. It is worth noting that the accuracy in the determination of the wavenumber depends on the grid size of numerical integration and is $\pm \Delta k/2$.

It is well known that a short Tollmien-Schlichting wave in a boundary layer can be excited by a long free-stream noise (Goldstein 1989). One of the key unresolved questions concerning the boundary-layer receptivity is the wavelength-adjustment mechanism. The Taylor-Couette problem is not the best model to explain this mechanism. However, we demonstrate here that a long stable wave can indeed excite a much shorter unstable wave through nonlinear wave interactions. Figure 6 shows the results of a numerical simulation in which a single mode k = 0.75 was given an amplitude density of 0.1 at time t = 0. The mode k = 0.75 is linearly stable, and decays to zero as it transfers energy to the mean flow. As indicated by Figure 6, nonlinear interactions excite the modes with wavenumbers 1.5 and 3. The mode k = 1.5, which is also linearly stable, initially grows and then decays to zero, while the mode k = 3, which is linearly unstable, grows and eventually reaches a supercritical finite-amplitude equilibrium state, together with its harmonic k = 6. The simulations were repeated using initial amplitude densities of 0.05 and 0.01. In all the cases, the final equilibrium state was the same. Numerical simulations starting with a single mode k = 1 and amplitude densities of 0.1, 0.05 and 0.01 also resulted in the same equilibrium state. In this case, nonlinear interactions generate the harmonics k = 2 and 3. The mode k = 2, although linearly unstable, does not grow to a finite-amplitude equilibrium as it loses all its energy due to nonlinear interactions, as might be anticipated from the results presented in Figure 3. The dominant mode in the final equilibrium state in this case is, however, k = 3, and not k = 4 as in figure 3. These results demonstrate that long wavelength disturbances can quickly excite modes that are linearly unstable and which grow to finite-amplitude equilibium states, although they are themselves linearly damped and eventually decay to zero. It is our opinion that the mechanism of nonlinear wave interaction is a credible mechanism for boundary-layer receptivity.

The selection of the equilibrium wavenumber is due to nonlinear wave interactions. The numerical results seem to suggest the following selection principle:

1) When the initial disturbance consists of a single dominant wave within the unstable region, the initial wave remains dominant in the final equilibrium state. Consequently, for a slowly accelerating cylinder, the critical wave is likely to be dominant.

2) When the initial condition consists of two waves with finite amplitudes in the unstable region, the final dominant wave is the one with the higher initial amplitude. If the two waves have the same initial finite amplitude, the dominant wave seems to be the one closer to the critical wave. On the other hand, if the initial amplitudes are very small, the faster growing wave becomes dominant.

3) When the initial disturbance is a uniform broad-band spectrum, the final dominant wave is the fastest linearly growing wave, if the initial amplitude is small. On the other hand, if the uniform noise level is not small, the critical wave is the dominant equilibrium one.

A similar selection principle has been found to hold in mixed-convection flow in a heated vertical annulus (Yao & Rogers, 1994). Since the final equilibrium state depends on the waveform of the initial disturbance, or equivalently, on ambient noise which cannot always be controlled, any property transported by the fluid can at best be determined within a limit of uncertainty associated with nonuniqueness. Since this nonuniqueness has been found to hold for two different flow situations, we believe that this uncertainty associated with nonuniqueness of the flow solution is generic to all fluid flows.

The torque, obtained by integrating the shear stress over the surface of the inner cylinder, is

$$G = 2\pi r_i^3 \Omega_1 \rho \nu \int_0^{2\pi/k_f} \left| \frac{\partial v(r_i, z)}{\partial r} - \frac{v(r_i, z)}{r_i} \right| dz.$$

The ratio of Taylor-vortex torque, $G_T$ and that for circular Couette flow, $G_C$ becomes

$$\frac{G_T}{G_C} = 1 - \frac{\eta(1+\eta)}{2}\{\Delta k A(k=0)\frac{\partial}{\partial r}\tilde{v}_1(k=0,r_i) + \Delta k^2 A^2(k_f)[\frac{\partial}{\partial r}\tilde{v}_2(k_f,-k_f,r_i) + c.c.]\},$$

where $k_f$ is the final equilibrium wavenumber. Values of $G_T/G_C$ for different wavenumbers $k_f$ are tabulated in Table 1 for R=88.1. The maximum value, occurring near the minimum critical wavenumber at the onset of instability, is 1.183; its minimum value is 1.096 at $k_f = 5$. The ratio $G_T/G_C$ is shown in Figure 7 as a function of the Reynolds number. The results obtained by Davey (1962) using the Stuart-Watson expansion for $k_C = 3.16$ are compared with the experimental data of Donnelly & Simon (1960). The the dashed upper bound is extrapolated from the present calculation at Re=88.1 and is corresponding the the critical wavenumber. Its values are slightly larger than the ones predicted by Davey. All possible torques predicted by the present study for different equilibrium states at R=88.1 is indicated by the vertical line. The variation in the torque between its maximum and minimum values, predicted by the present study, is about ten percent at Re=88.1. Since the equilibrium state is not unique and depends on the initial conditions which usually cannot be controlled in engineering systems, an uncertainty, associated with the nonuniqueness, should be considered when any accurately measured value of Taylor-vortex torque is used in practice.

## 4. Conclusion

The current study confirms that the supercritical equilibrium state of Taylor-Couette instability, slightly above the critical Taylor number, depends on the initial disturbance, as observed experimentally, and that it is not unique. Extrapolating this fact to turbulence, it is our opinion, since the *time* average will depend on the initial condition, it will not equal to the *ensemble* average even for stationary turbulence. From an application point of view, only time average has physical signifiance. The results also demonstrate that a long stable wave can quickly excite a much shorter unstable wave. This observation suggests that nonlinear wave interaction may be a credible mechanism for boundary-layer receptivity. The integrodifferential equation, (12), contains considerable information. In this paper, this equation was only lightly probed. The current formulation extended to travelling wave states will be reported later (Yao and Ghosh Moulic 1993).

Yao, L. S., and Ghosh Moulic, S., 1994, "Uncertainty of convection," To appear in Internatioonal Journal of Heat & Mass Transfer

**Appendix A. Amplitude equations for monochromatic waves**

The numerical results of (12) indicate that the energy transfers nonlinearly among the mean flow, the donminant wave and its harmonics. The modification of the mean flow is at same order of magnitude of the dominant wave. If there is no energy transfer to waves other than the selected disturbance wave, a set of simpler amplitude equations can be derived to replace equation (12) of a continuous spectrum. Rigorously, the expansion should include the amplitude density functions for the mean flow, the dominant disturbance wave and its harmonics. We assume that the harmonics are negligiblely small to simplify the algebra in order to demonstrate the concept clearly. Substitution of the amplitude density function,,

$$A(k, t) = \left[ A_0 \delta(k) + A_0^* \delta(k) \right] + \left[ A_{k_0} \delta(k - k_0) + A_{k_0}^* \delta(k + k_0) \right], \tag{A.1}$$

into equation (12) results

$$\frac{dA_{k_0}}{dt} = a(k_0) A_{k_0} + \left[ a_{31} A_0 A_{k_0} \right] + \left( a_{41} |A_{k_0}|^2 + a_{42} |A_0|^2 \right) A_{k_0}, \tag{A.2}$$

and

$$\frac{dA_0}{dt} = a(0) A_0 + \left( b_{31} |A_0|^2 + b_{32} |A_{k_0}|^2 \right) + \left( b_{41} |A_{k_0}|^2 + b_{42} |A_0|^2 \right) A_0, \tag{A.3}$$

where $a_{ij}$ and $b_{ij}$ are interacting constants. $A_o$ is the amplitude function for the deformation of the mean flow and $A_{k_0}$ is for the disturbance wave. The terms inside of first square bracket on the right-hand side of (A.2) and (A.3) are from three-wave resonance. The terms in the second square bracket are from four-wave resonance. The terms multiplied by $a_{31}$ and $a_{42}$ represent the nonlinear energy transfer between the mean flow and the disturbance wave, and are not included in the Landau and Ginzburg-Landau equations. The term multiplied by $a_{41}$ represnts the self-interaction of the disturbance wave as explained in equation (18), and is the only nonlinear term in the Landau and Ginzburg-Landau equations.

There are two types of nonlinear interaction between two waves. One is direct interaction, if they satisfy the resonance condition, which has been pointed out by Landau. The other is *indirect* interaction in which the two waves interact simultaneously with the mean flow (k=0) as indicated in (A.2) and (A.3). The indirect interaction always exists and has so far been overlooked. It is clear that only linear energy transfer, the terms multiplied by the growth rate, $a$, is considered in the Landau and Ginzburg-Landau equations. Without the terms of nonlinear energy transfer between the mean flow and the disturbance wave, they completely ignore the indirect wave interaction. Equations similar to (A.2) and (A.3) have been derived for the interaction of two disturbance waves, but none of them have considered an additional equation for the deformation of the mean flow.

Table 1. Variation of torque with wavenumber

| $k_f$ | 2.75 | 3.0 | 3.25 | 3.5 | 3.75 | 4.0 | 4.25 | 4.5 | 4.75 | 5.00 |
|---|---|---|---|---|---|---|---|---|---|---|
| $G_T/G_c$ | 1.150 | 1.172 | 1.181 | 1.183 | 1.179 | 1.172 | 1.161 | 1.144 | 1.122 | 1.096 |

**List of Figures.**

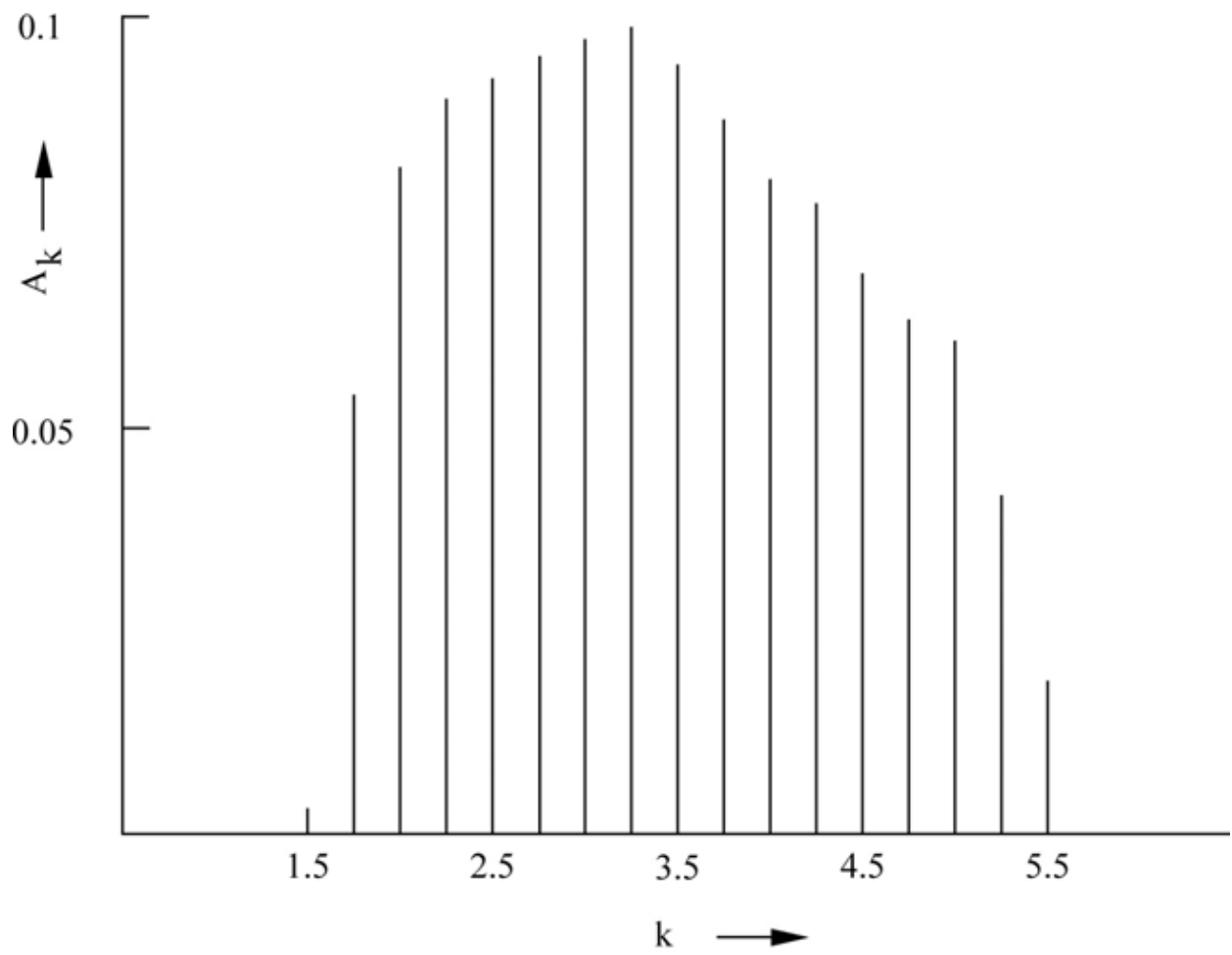

**Figure 1**
Equilibrium amplitudes of Taylor vortex flows with different wavenumbers for discrete monochromatic waves

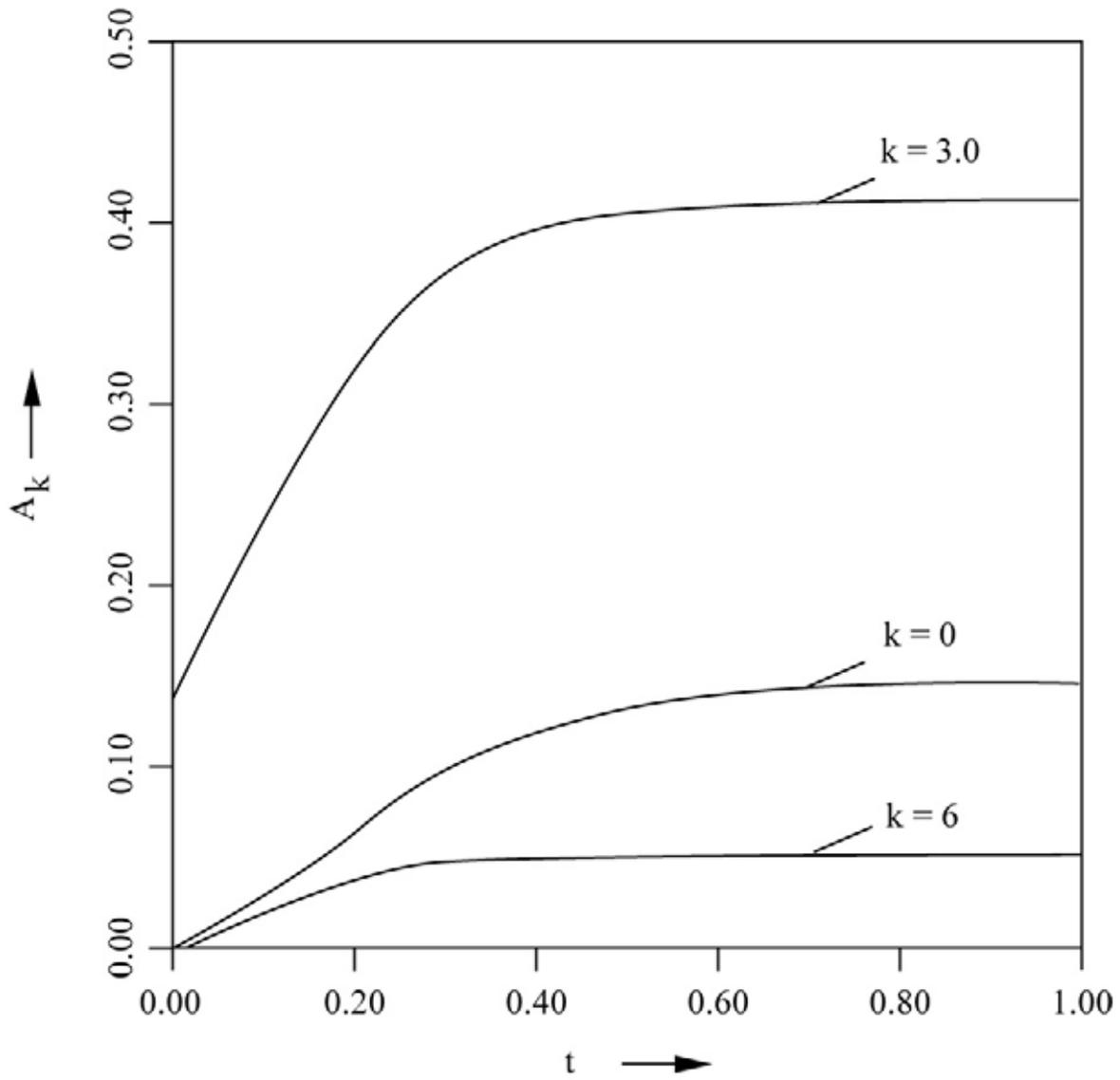

**Figure 2 (a)**
Evolution of the amplitude density function for a single initial mode with wavenumber k = 3.

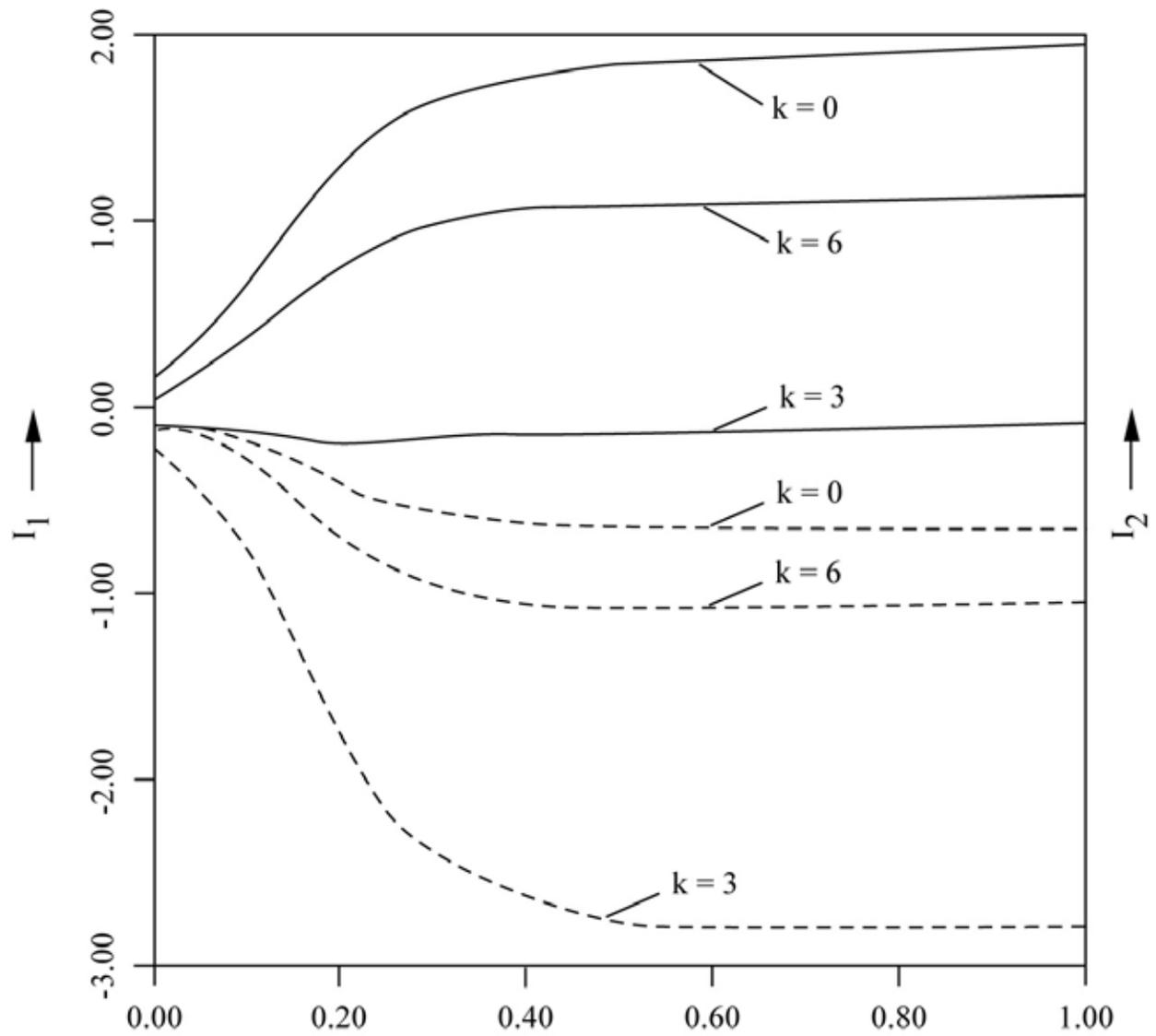

**Figure 2 (b)**
Evolution of the interaction integrals $I_1$ and $I_2$ for a single initial mode with wavenumber $k = 3$. The solid lines indicate the integal $I_1$, while the dotted lines refer to the integral $I_2$.

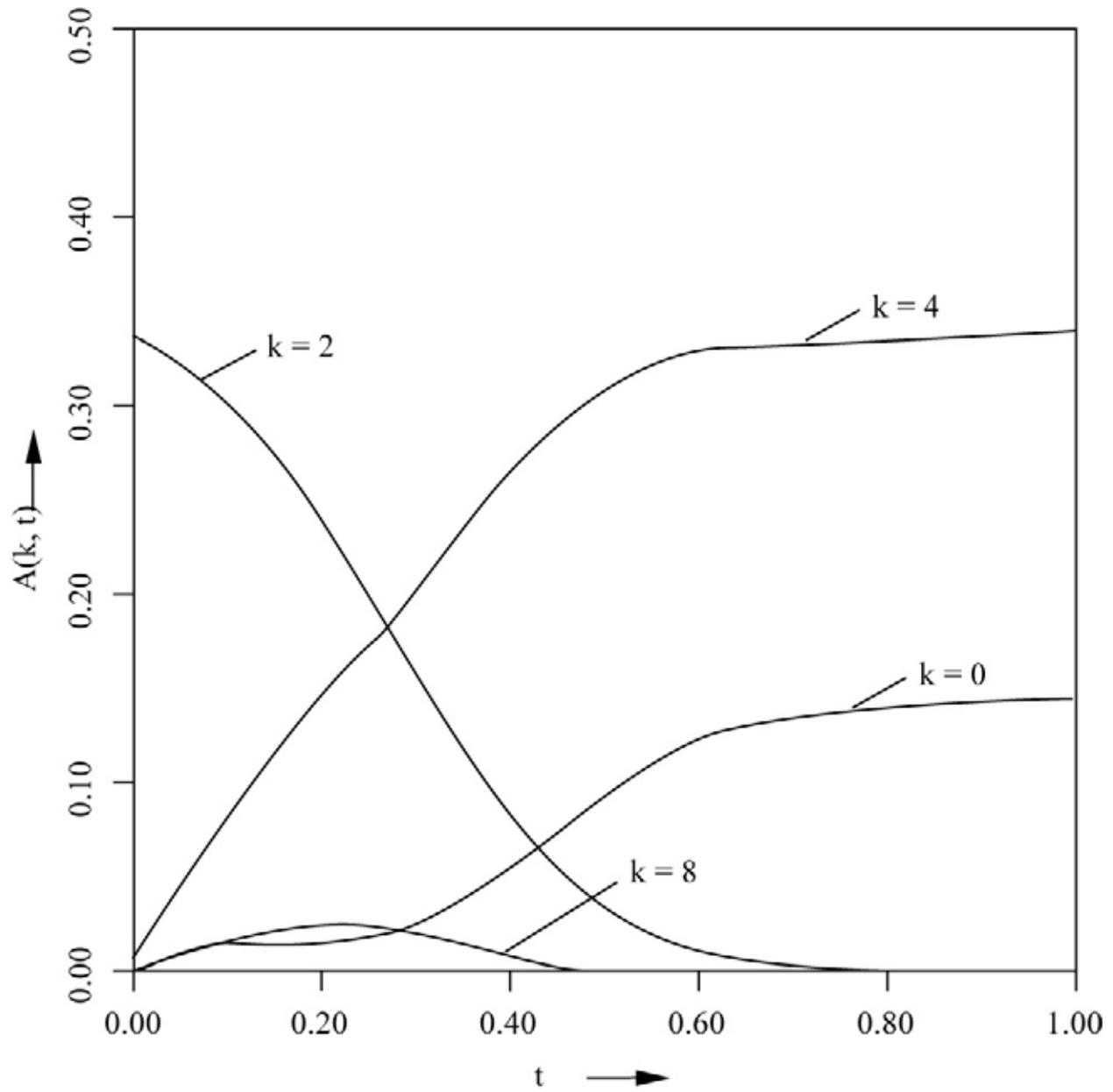

**Figure 3 (a)**
Evolution of the amplitude density function for a single initial mode with wavenumber k = 2, illustrating sideband instability.

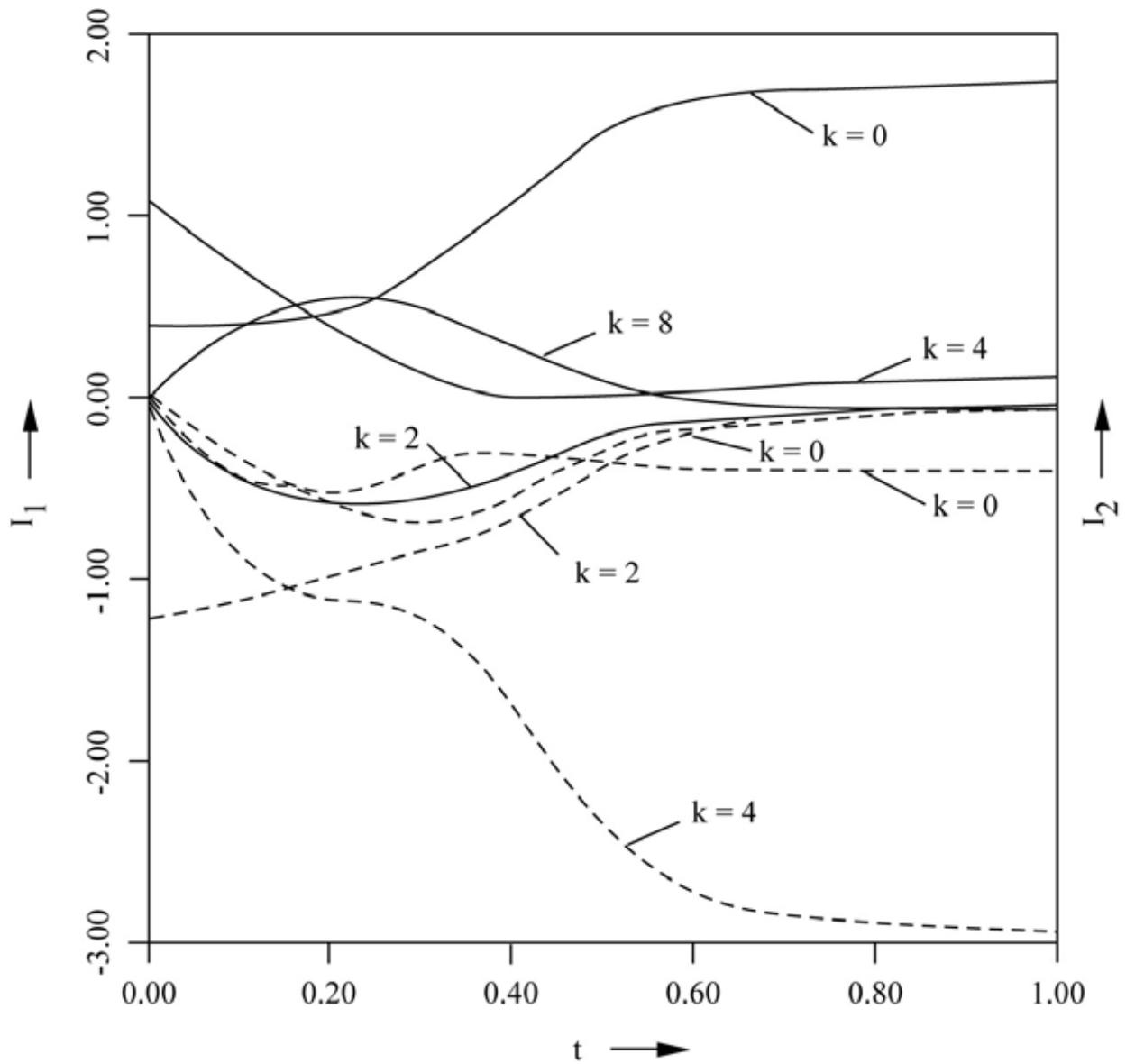

**Figure 3 (b)**
Evolution of the interaction integrals $I_1$ and $I_2$ for a single initial mode with wavenumber k = 2. The solid lines indicate the integal $I_1$, while the dotted lines refer to the integral $I_2$.

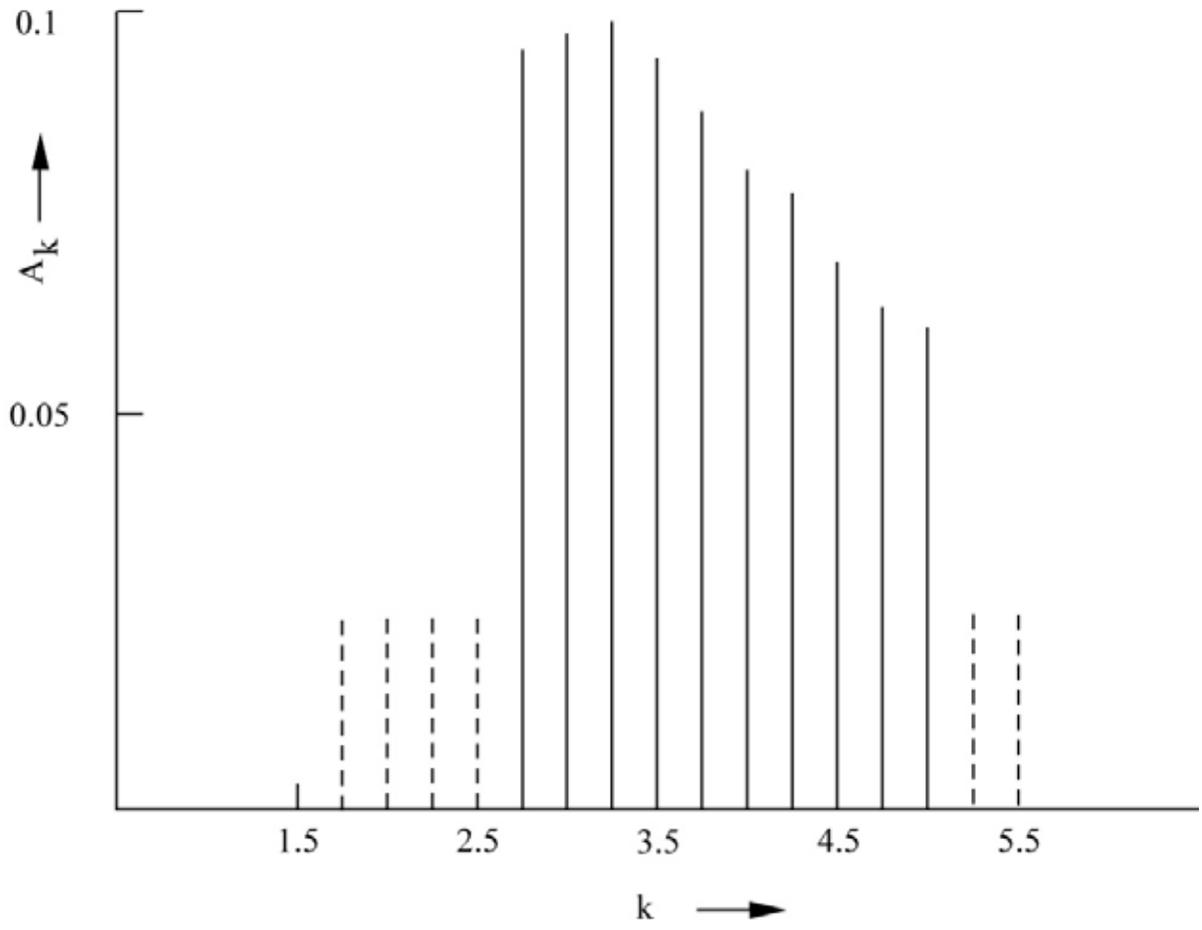

**Figure 4**
Equilibrium amplitudes of dominant modes obtained with a single initial dominant mode.

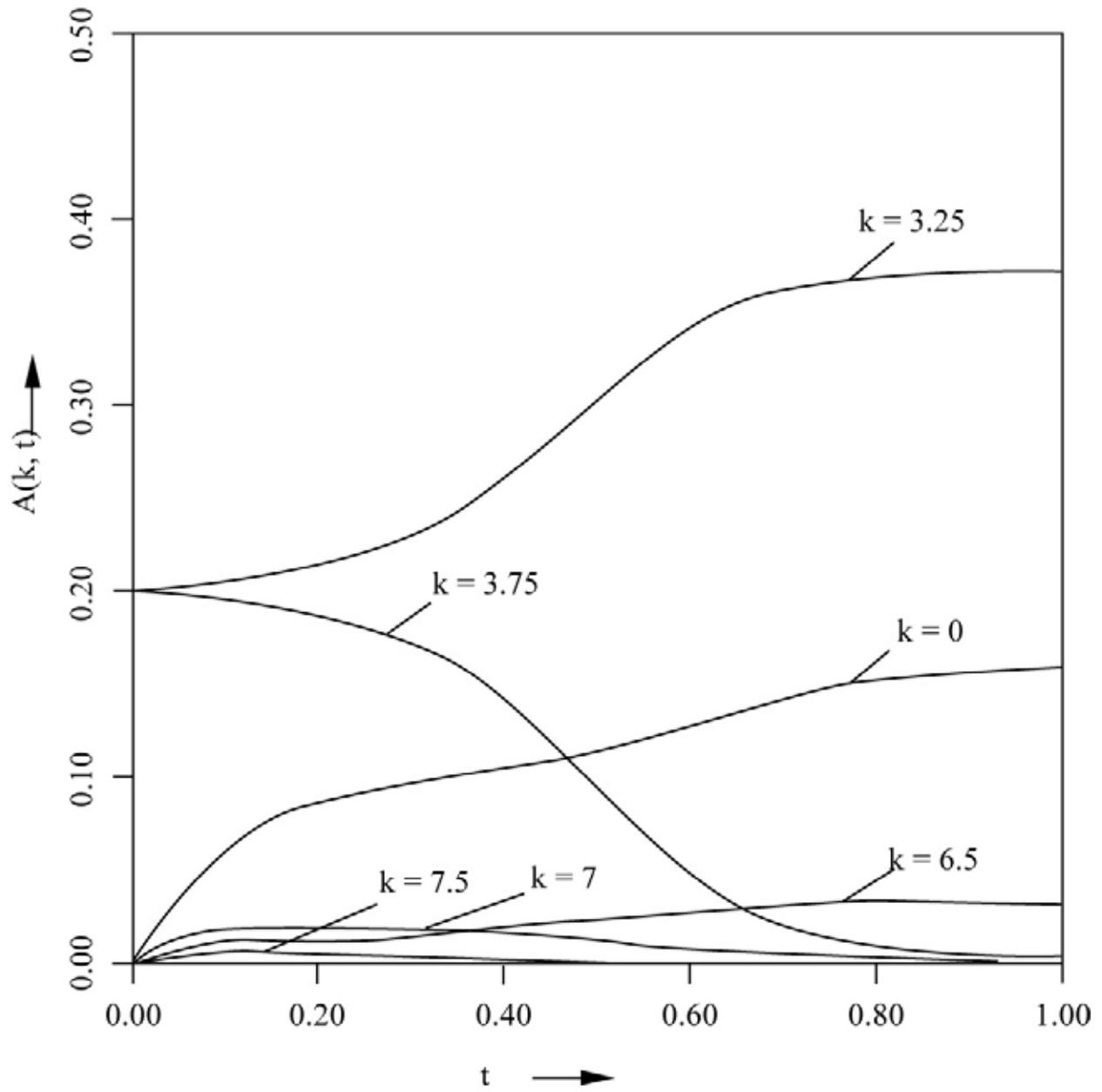

**Figure 5 (a)**
Evolution of the amplitude density function for a pair of initial modes with wavenumbers k = 3.25 and k = 3.75 and initial amplitudes of 0.2.

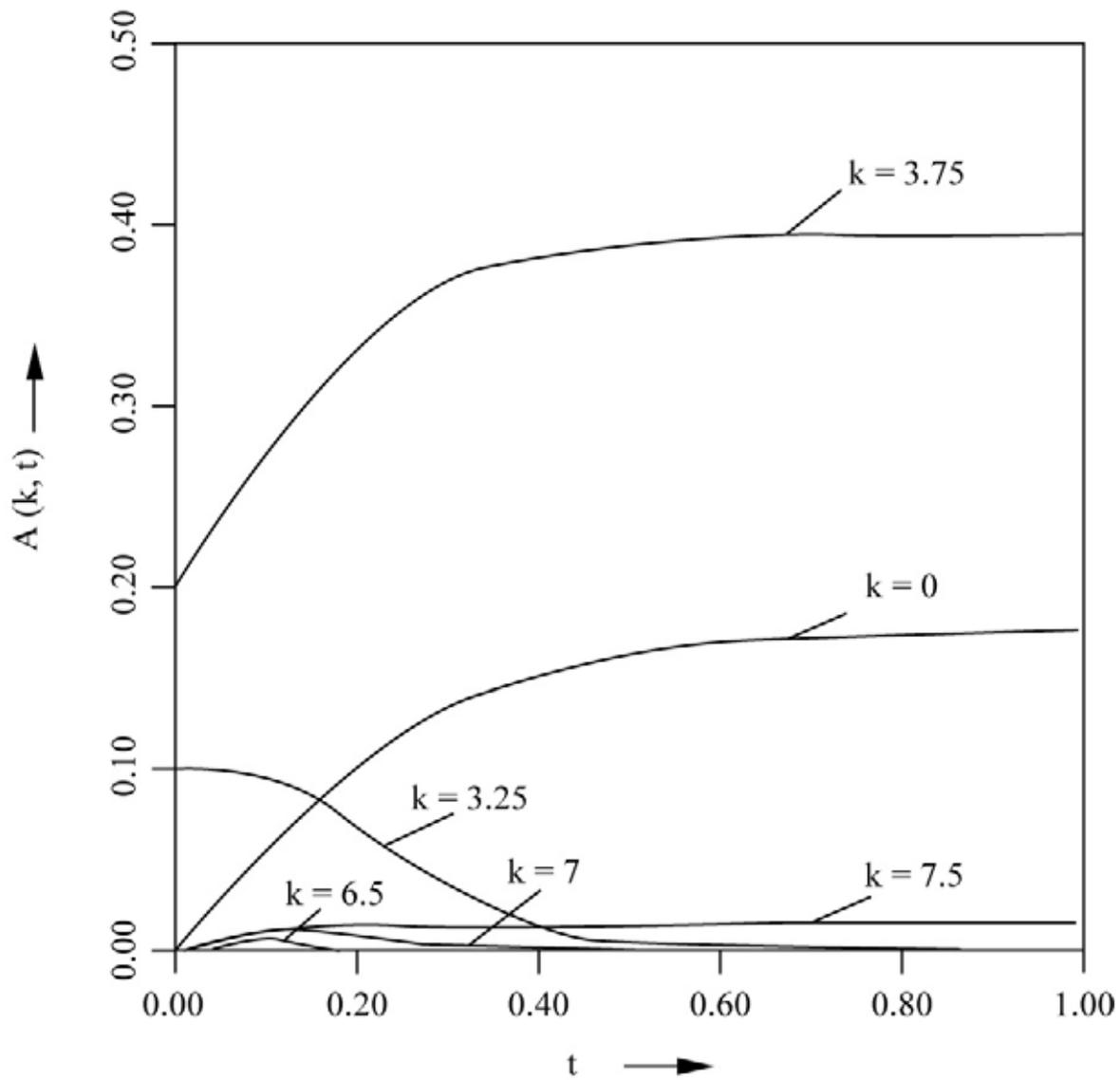

**Figure 5 (b)**
Evolution of the amplitude density function for a pair of initial modes with wavenumbers k = 3.25 and k = 3.75 and amplitudes 0.2 and 0.1 respectively.

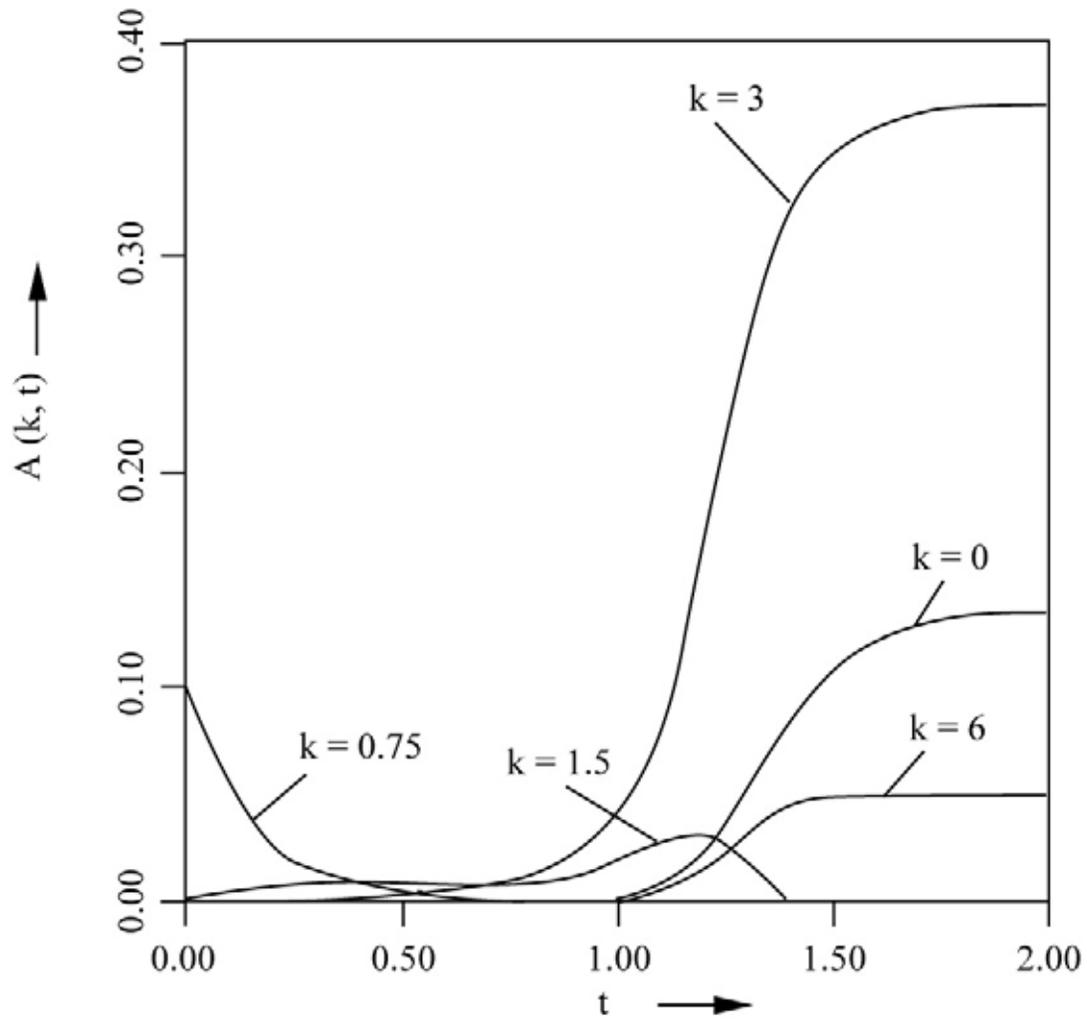

**Figure 6**
Evolution of the amplitude density function for a single linearly stable initial mode with wavenumber k = 0.75.

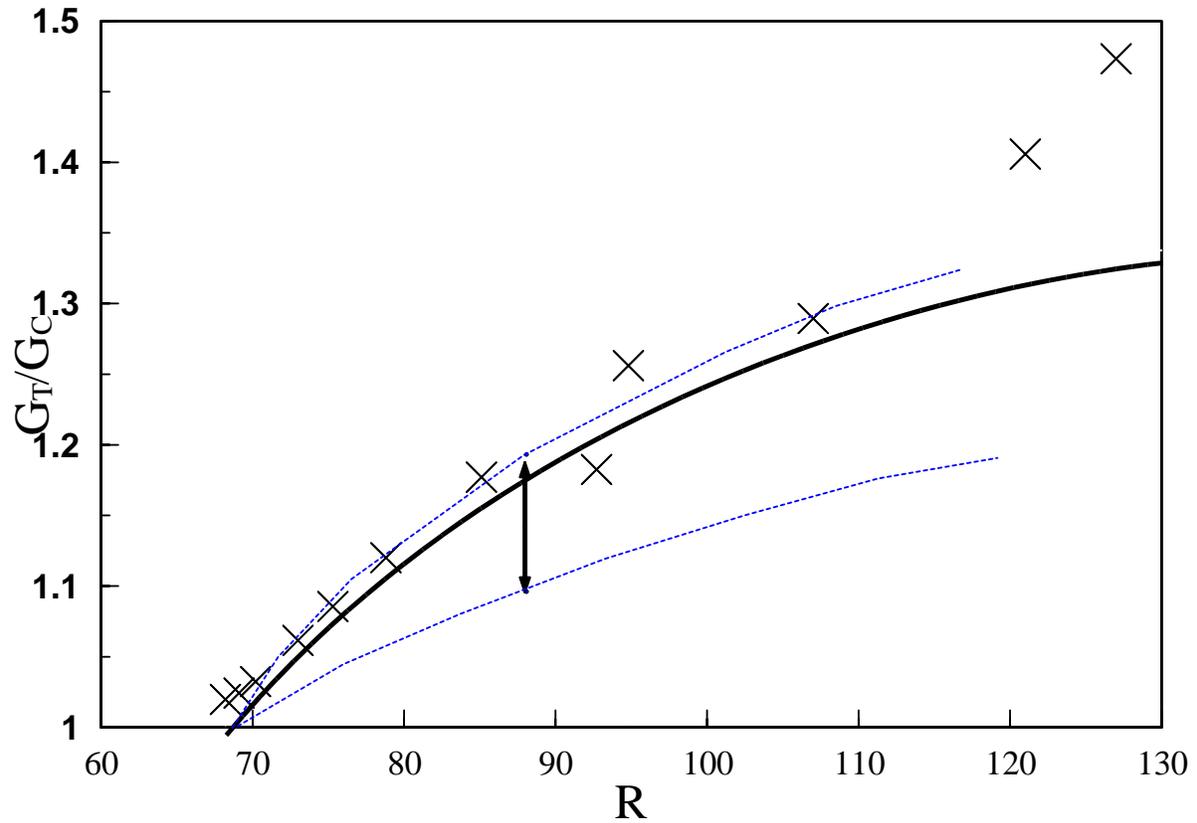

Figure 7. Ratio of the torques of Taylor-vortex flow and Couette-flow as functions of the Reynolds number. The symbol x represents the experimental data of Donnelly and Simon (1960), the curve is the prediction of Davey (1962), and the vertical line represents all the possible equilibrium states predicted by the present analysis.